\documentclass[]{spie}  

\usepackage{amsmath,amsfonts,amssymb}
\usepackage{graphicx}
\usepackage[colorlinks=true, allcolors=blue]{hyperref}
\usepackage{siunitx}

\title{The space coronagraph optical bench (SCoOB): 2. wavefront sensing and control in a vacuum-compatible coronagraph testbed for spaceborne high-contrast imaging technology}

\author[a]{Kyle Van Gorkom}
\author[a]{Ewan S. Douglas}
\author[b]{Jaren N. Ashcraft}
\author[a]{Sebastiaan Haffert}
\author[a,b]{Daewook Kim}
\author[b]{Heejoo Choi}
\author[a]{Ramya M. Anche}
\author[a]{Jared R. Males}
\author[b]{Kian Milani}
\author[b]{Kevin Derby}
\author[a]{Lori Harrison}
\author[a]{Olivier Durney}

\affil[a]{Department of Astronomy and Steward Observatory, University of Arizona, 933 N. Cherry Ave., Tucson, AZ 85719, USA}
\affil[b]{James C. Wyant College of Optical Sciences, University of Arizona, Meinel Building 1630 E. University Blvd., Tucson, AZ. 85721}
\authorinfo{Further author information: (Send correspondence to K.V.G)\\K.V.G.: E-mail: kvangorkom@arizona.edu}

\begin{document} 
\maketitle

\begin{abstract}

The 2020 Decadal Survey on Astronomy and Astrophysics endorsed space-based high contrast imaging for the detection and characterization of habitable exoplanets as a key priority for the upcoming decade. To advance the maturity of starlight suppression techniques in a space-like environment, we are developing the Space Coronagraph Optical Bench (SCoOB) at the University of Arizona, a  new thermal vacuum (TVAC) testbed based on the Coronagraphic Debris Exoplanet Exploring Payload (CDEEP), a SmallSat mission concept for high contrast imaging of circumstellar disks in scattered light. When completed, the testbed will combine a vector vortex coronagraph (VVC) with a Kilo-C microelectromechanical systems (MEMS) deformable mirror from Boston Micromachines Corp (BMC) and a self-coherent camera (SCC) with a goal of raw contrast surpassing $10^{-8}$ at visible wavelengths. In this proceedings, we report on our wavefront sensing and control efforts on this testbed in air, including the as-built performance of the optical system and the implementation of algorithms for focal-plane wavefront control and digging dark holes (regions of high contrast in the focal plane) using electric field conjugation (EFC) and related algorithms.

\end{abstract}


\section{INTRODUCTION}
\label{sec:intro}  

Future space observatories tasked with the goal of detecting and characterizing earth-like exoplanets will require coronagraphs and wavefront control algorithms capable of achieving and maintaining contrasts on the order of $10^{-10}$. To aid in the maturation program for high contrast imaging technology, we are developing the space coronagraph optical bench (SCoOB) at the University of Arizona Space Astrophysics Lab (UASAL), a thermal vacuum high contrast imaging testbed designed to enable demonstrations of starlight suppression techniques in a space-like environment. In this report, we describe our initial wavefront sensing and control efforts on the testbed.

In Section \ref{sec:intro_testbed}, we give an overview of the testbed. Sections \ref{sec:strehl} and \ref{sec:dh} describe our focal-plane wavefront control strategies and results for Strehl ratio optimization and digging dark holes for high contrast imaging. Finally, in Section \ref{sec:future}, we outline our plans for future upgrades to the testbed.

\section{Testbed overview}
\label{sec:intro_testbed}

 A brief overview of the key features of the testbed is given here. Details of the initial optical design are given in  Maier et al.\ 2020\cite{maier} and the updated design, assembly, and initial as-built performance are described in Ashcraft et al.\ 2022\cite{jaren}.

\textbf{Deformable mirror}: The system DM is a Kilo-C 1.5$\si{\micro m}$-stroke MEMS device from BMC with 952 actuators and one non-responsive actuator. The beam footprint at the DM plane is approximately 9.6mm for 32 actuators across the pupil.

\textbf{Focal plane masks}: An image plane in an f/47 beam downstream of the DM is available for focal plane masks (FPMs). Initial experiments on the testbed have primarily made use of a knife-edge FPM. A triple-grating charge-6 vector vortex coronagraph (TGVVC)\cite{doelman} is currently installed for testing. Some initial lab measurements with the TGVVC are presented in Doelman et al. 2022\cite{doelmanspie}. Additional VVC \cite{Foo, mawet} devices from BEAM Co are being procured.

\textbf{Lyot stop}: A $\sim$95\% Lyot stop (LS) is installed for starlight suppression in concert with the FPM. The Lyot stop includes a pinhole outside the geometric pupil for self-coherent camera (SCC)\cite{galicher} experiments. A tolerancing analysis of the SCC is presented in Derby et al.\ 2022\cite{derby}.

\textbf{Cameras}: The focal-plane science camera is an ASI294MM Pro CMOS device with 2.32$\si{\micro m}$ pixels and 1.2-7.3e$^{-}$ read noise in an f/18 beam. A pellicle downstream of the LS relays the beam to a Basler acA2500-60um camera in a pupil plane.

\textbf{Source}: The current source is a PD-LD SLM-632.8 frequency-stabilized laser diode that outputs 22mW of power.

\textbf{Software}: The \texttt{cacao}\cite{cacao} package forms the basis of the testbed control software, with additional packages inherited from MagAO-X\cite{magaox}. Most closed-loop control operations are performed via Python interfaces with shared memory images in \texttt{cacao}.

The testbed is currently operated in-air on a pneumatic optical table but will be moved into a thermal vacuum (TVAC) chamber later in 2022.

\begin{figure} [ht]
   \begin{center}
   \begin{tabular}{c}
   \includegraphics[height=7cm]{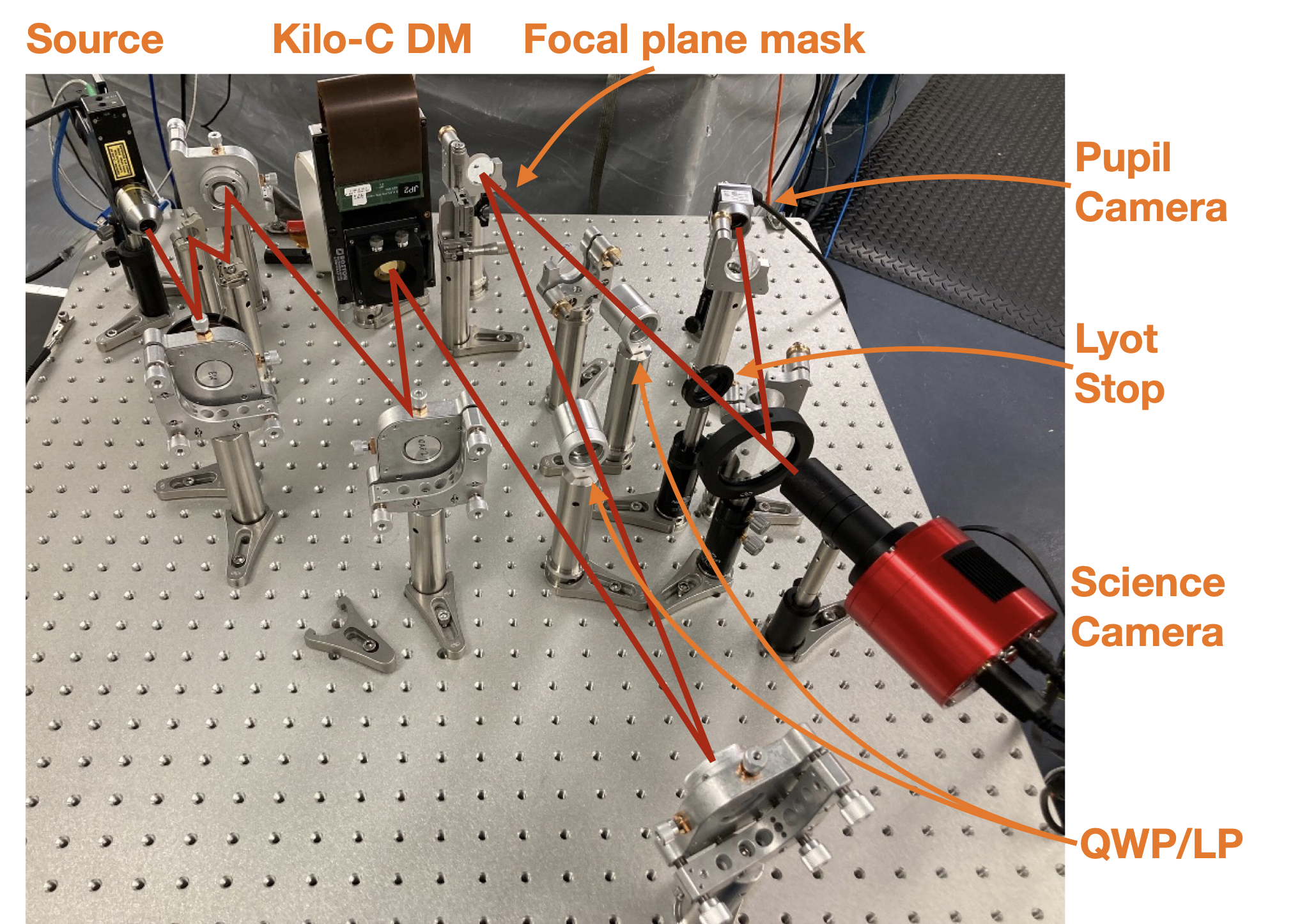}
   \end{tabular}
   \end{center}
   \caption[example] 
   { \label{fig:layout} 
Image of SCoOB, with key elements labelled. Refer to Ashcraft et al.\ 2022\cite{jaren} for the optical design.}
\end{figure} 

\section{Strehl ratio optimization}
\label{sec:strehl}

Prior to digging any dark holes, we employ a phase retrieval (PR) algorithm in closed-loop with the DM to drive out phase aberrations and maximize the testbed Strehl ratio.

The voltage bias that maximizes the inter-actuator stroke on a Kilo-C device is found around 70\%; however, provided that the commands required to clean up the wavefront and dig a dark hole are not limited by stroke, it may be more optimal to take advantage of the approximately quadratic stroke-voltage relationship and choose a smaller operating bias to lower the contrast floor that arises from discretization effects. Following the analysis in Ruane et al. 2020 \cite{ruane}, we estimate the contrast floor due to our 14-bit COTS drive electronics and Kilo-C DM. We apply an extra factor of 2 to account for the discrepancy found between the analytical and measured contrast floor. With a gain of 8.7 nm/V at 70\% bias, we expect a floor of $\sim 10^{-8}$; at 40\%, our device has a gain of 2.9 nm/V for a contrast floor of $\sim 10^{-9}$. We choose this latter bias around which to flatten the DM.

\begin{figure} [ht]
   \begin{center}
   \begin{tabular}{c}
   \includegraphics[height=8cm]{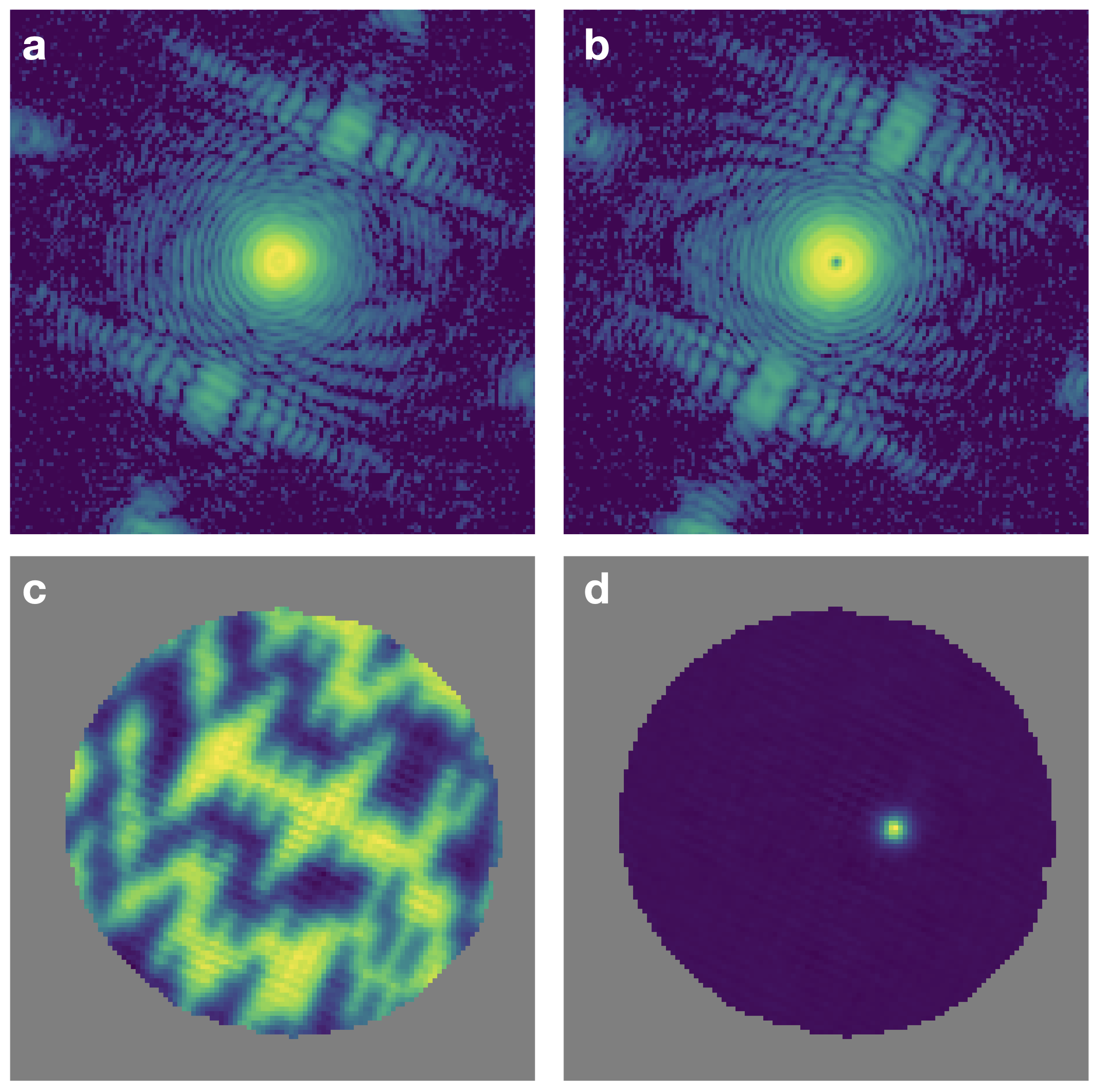}
   \end{tabular}
   \end{center}
   \caption[example] 
   { \label{fig:pr_examples} 
Examples of (a) and (b) PSF measurements with defocus diversity induced by the DM shown here in log scale, (c) a PR estimate of the pupil phase map corresponding to a single Hadamard mode, and (d) pupil phase map of an actuator influence function reconstructed from Hadamard modes.}
\end{figure}

We employ a parametric phase retrieval algorithm adapted from Thurman et al. 2009\cite{thurman} and implemented on MagAO-X \cite{memagaox} but modified to replace the angular spectrum propagation to defocused planes with a pair of defocus probes on the DM. We calibrate the PR response matrix similar to the procedure described in Van Gorkom et al. 2021 \cite{memagaox} but employ a set of Hadamard probes on the DM to calibrate the system \cite{kasper} in place of grid patterns. Actuators outside the beam footprint on the DM are removed from the calculated control matrix and commanded to the mean of nearest-neighbor actuators. Examples of the measured probed PSFs, phase estimates of the Hadamard modes, and reconstructed influence functions are shown in Figure \ref{fig:pr_examples}.

\begin{figure} [ht]
   \begin{center}
   \begin{tabular}{c}
   \includegraphics[height=4cm]{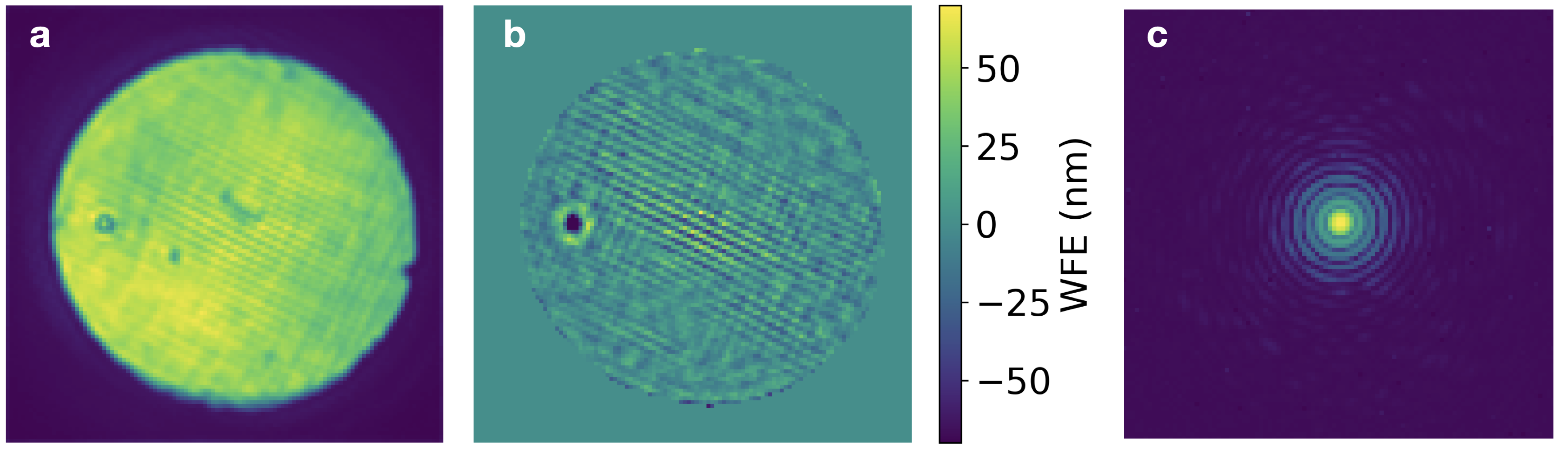}
   \end{tabular}
   \end{center}
   \caption[example] 
   { \label{fig:wfe_psf} 
Examples of the retrieved pupil-plane (a) amplitude and (b) phase after running PR in closed loop, and (c) the measured PSF, shown in log scale. The dominant feature in the phase map is the bad actuator on the Kilo-C DM.}
\end{figure}

A typical example of a retrieved phase map after closed loop wavefront optimization is shown in Figure \ref{fig:wfe_psf}, along with a measurement of the optimized PSF. We estimate a residual wavefront of approximately 10 nm RMS, and a Strehl ratio of 0.95. The amplitude of our defocus diversity is relatively small  ($\pm 1$ wave) and limited by the DM stroke available from the 40\% voltage bias, which suggests our PR has reduced sensitivity to high-order aberrations \cite{dean}, likely resulting in somewhat optimistic estimates of the residual wavefront RMS and Strehl ratio. 

\section{Wavefront control for high contrast imaging}
\label{sec:dh}

Initial demonstrations of high contrast imaging on the testbed employed a knife-edge FPM combined with a $\sim 95\%$ Lyot stop and a HeNe laser at $\lambda=633$ nm. The knife-edge was aligned to block out to 1-2 $\lambda/D$. Since we have only a single DM in the system, we dig only one-sided DHs. 

 To dig a dark hole without the need to develop an end-to-end optical model of the testbed, we turn to implicit electric field conjugation (I-EFC)\cite{haffert2022magaox, iefc}. The key insight of I-EFC is that the intensity in the dark zone can be controlled from probed focal-plane measurements without the need to explicitly estimate the electric field. A brief and incomplete overview of the algorithm is provided here. Readers interested in a more complete treatment of the algorithm should refer to Haffert et al. 2022\cite{haffert2022magaox} and Haffert et al. (in prep) \cite{iefc}.

In the small aberration regime, the pupil plane field can be expressed as:
\begin{equation}
	E_\mathrm{pupil} = A_{ab} e^{i \phi_{ab}} e^{i \phi_{DM}} \approx A_{ab} [1 + i \phi_{ab} + i \phi_{DM} ] = E_0 + E_{DM},
\end{equation}
where $E_0 = A_{ab} [1 + i \phi_{ab}]$ and $E_{DM} = i A_{ab} \phi_{DM}$. In the focal plane, the electric field is given by
\begin{equation}
	E_\mathrm{focal} = 	\hat{E}_0 + \hat{E}_{DM},
\end{equation}
where $\hat{x} = \mathcal{C}\{x\}$ and $\mathcal{C}$ is an operator that propagates the electric field from the pupil plane to focal plane. The intensity is given by
\begin{equation}
	I_\mathrm{focal} = | E_\mathrm{focal} |^2 = |\hat{E}_0|^2 + |\hat{E}_{DM}|^2 + 2 \mathcal{R}\{ \hat{E}_{DM}^* \hat{E}_{0} \}	
\end{equation}

In the pairwise probing (PWP) technique employed in classical EFC\cite{efc}, we choose a pair of positive and negative probes $\hat{E}_{DM} = \pm \hat{E}_{p}$ and take the difference of the intensity measurements to get
\begin{equation}\label{eqn:iefc_di} 
	\Delta I = 4 \mathcal{R} \{  \hat{E}_{p}^* \hat{E}_0 \}.	
\end{equation}

With a set of $K$ probes, we can cast Equation \ref{eqn:iefc_di} into matrix form:
\begin{equation}
 \phi \equiv \frac{1}{4} \begin{bmatrix}
 	\Delta I_1 \\ \vdots \\ \Delta I_K
 \end{bmatrix}
 = \begin{bmatrix}
 	\mathcal{R}\{ \hat{E}_p^1 \} & \mathcal{I}\{ \hat{E}_p^1 \} \\ \vdots & \vdots \\ \mathcal{R}\{ \hat{E}_p^K \} &\mathcal{I}\{ \hat{E}_p^K \}
 \end{bmatrix}
 \begin{bmatrix}
 	\mathcal{R} \{ \hat{E}_0 \} \\ \mathcal{I} \{ \hat{E}_0 \}
 \end{bmatrix}
\end{equation}

In I-EFC, we choose two pairs of positive and negative probes such that $\hat{E}_{DM} = \pm \hat{E}_{p} \pm \hat{E}_{c}$ and take the double difference to get
\begin{equation}\label{eqn:iefc_ddi}
	\Delta \Delta I = 8 \mathcal{R} 	\{ \hat{E}_{p}^* \hat{E}_{c} \}.
\end{equation}

And with a set of $N$ modal calibration probes, we can extend  Equation \ref{eqn:iefc_ddi} to construct an empirical response matrix for the probed field:
\begin{equation}
	\frac{1}{8} \begin{bmatrix}
 	\Delta \Delta I_1^1 & \cdots & \Delta\Delta_1^N \\ \vdots & \ddots & \vdots \\ \Delta\Delta I_K^1 & \cdots & \Delta\Delta I_K^N
 \end{bmatrix} = \begin{bmatrix}
 	\mathcal{R}\{ \hat{E}_p^1 \} & \mathcal{I}\{ \hat{E}_p^1 \} \\ \vdots & \vdots \\ \mathcal{R}\{ \hat{E}_p^K \} &\mathcal{I}\{ \hat{E}_p^K \}
 	\end{bmatrix}
 \begin{bmatrix}
 	\mathcal{R}\{ \hat{E}_c^1 \} & \cdots & \mathcal{R}\{ \hat{E}_c^N \} \\
 	\mathcal{I}\{ \hat{E}_c^1 \} & \cdots & \mathcal{I}\{ \hat{E}_c^N \} 
 \end{bmatrix} = G u,
\end{equation}
where $u$ is the vector of $N$ DM modes. Then the I-EFC control law can be found by solving

\begin{equation}
	u = \underset{u}{\mathrm{argmin}} \bigg\{ |\phi + Gu|^2 + \lambda|u|^2 \bigg\}.
\end{equation}

For the I-EFC experiments conducted on SCoOB thus far, we typically choose $K=3$ probes of random DM commands generated from power law PSD filtered to maximize the signal over a particular range of $\lambda/D$ separations and then build the response matrix with $N=1024$ Hadamard calibration modes. We select the dark hole by applying a weighting to the focal-plane measurements and smooth it to handle the boundaries of the dark hole. Actuators outside the beam footprint on the DM are filtered from the control and assigned a value from the mean of their nearest neighbors. The regularization $\lambda$ is chosen heuristically, typically starting with a large value and decreasing as the dark hole contrast deepens. Future efforts on the testbed will be directed at optimizing these choices.

\begin{figure} [ht]
   \begin{center}
   \begin{tabular}{c}
   \includegraphics[height=8cm]{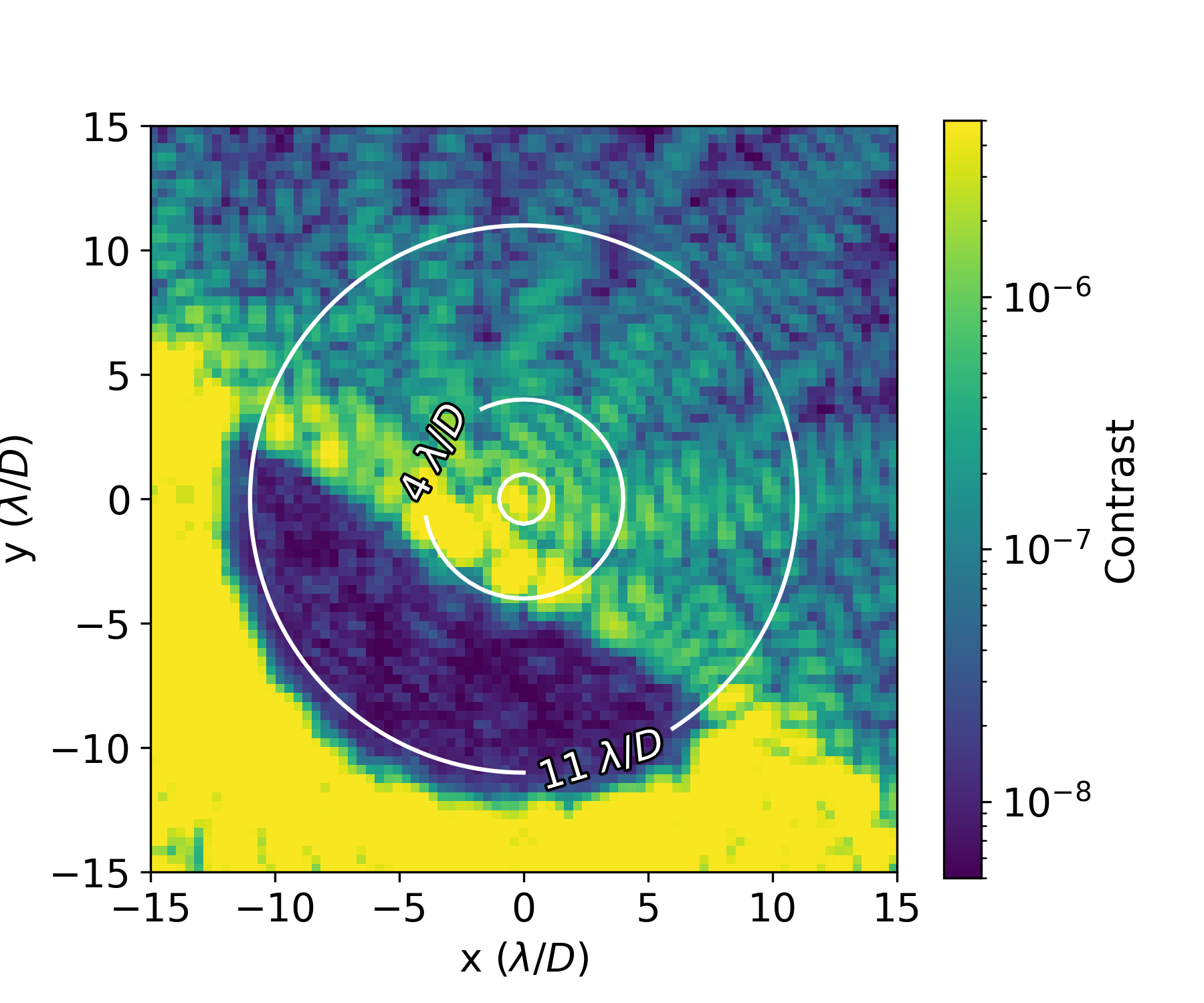}
   \end{tabular}
   \end{center}
   \caption[example] 
   { \label{fig:knife_dh} 
Dark hole dug with a knife-edge FPM and I-EFC, achieving a mean contrast of $2.3\times10^{-8}$ from $4-11\lambda/D$ in monochromatic light at 633nm.}
\end{figure} 

The best monochromatic contrast achieved to-date on SCoOB with a knife-edge and Lyot stop is shown in \ref{fig:knife_dh}, with a mean contrast of $2.3\times10^{-8}$ from $4-11\lambda/D$. The current testbed contrast is likely limited by some combination of atmospheric turbulence, mechanical disturbances, laser stability, scattered light, and polarization aberrations. Many upgrades to the system are planned to mitigate these effects, described in the next section.

\section{Conclusions and future work}
\label{sec:future}

SCoOB is a high contrast imaging testbed at the University of Arizona that has demonstrated a $2.3 \times 10^{-8}$ contrast in monochromatic light under unstable laboratory conditions. A number of significant upgrades are underway and slated to be completed by the end of 2022, which will enable the demonstration of broadband dark holes in vacuum under thermal control. These efforts include improvements in the following areas:

\textbf{Source upgrades}: We plan to replace our current laser diode with a supercontinuum fiber laser (the FYLA Iceblink) to demonstrate dark hole digging in 1 to 20\% bandpasses with central wavelengths ranging from 450 to 700 nm. A portion of this effort will be directed at reworking the current source optics and spatial filter for a cleaner input beam.

\textbf{Focal plane masks}: We recently integrated a triple-grating charge-6 VVC \cite{doelman} with our testbed and expect to receive a charge-6 VVC from BEAM Co in the late summer. The majority of our near-future wavefront control efforts will make use of these devices.

\textbf{Focal plane wavefront control strategies}:  In addition to our ongoing efforts to refine our I-EFC implementation, an end-to-end Fresnel model has been developed to enable classical EFC on the testbed. Dark hole digging with the SCC will also be demonstrated in the upcoming months.

\textbf{Low-order wavefront sensors}: Plans include testing low-order wavefront sensing approaches including a reflective Lyot stop\cite{singh, mendillo} and a Zernike wavefront sensor\cite{ndiaye}.

\textbf{Scattered light}: A field stop will be added to a focal plane between the Lyot stop and the science plane to limit scattered light at the science plane\cite{hcst,hcst2}.

\textbf{Thermal vacuum chamber}: In order to increase stability and eliminate atmospheric effects, we plan to integrate the testbed with a ``TVAC'' chamber from Rydberg Vacuum Sciences in Fall 2022. We are in the process of acquiring vacuum-compatible cameras, DM cables, and a fine-steering mirror to aid in aligning the PSF to the FPM.

\begin{figure} [ht]
   \begin{center}
   \begin{tabular}{c}
   \includegraphics[height=6cm]{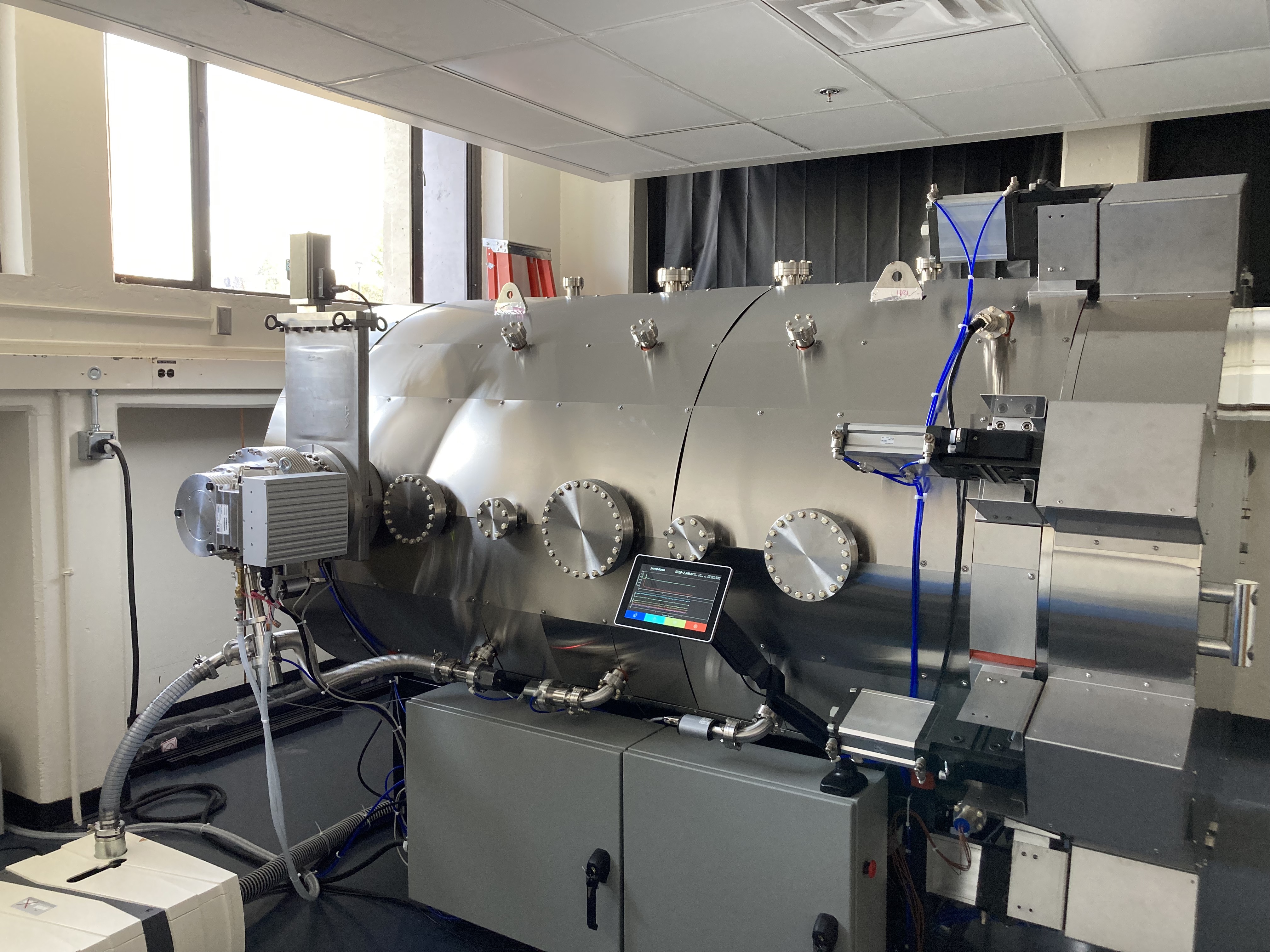}
   \end{tabular}
   \end{center}
   \caption[example] 
   { \label{fig:tvac} 
The thermal vacuum chamber at UArizona.}
\end{figure} 

\appendix    

\acknowledgments 
Portions of this work were supported by the Arizona Board of Regents Technology Research Initiative Fund (TRIF).
S.H.'s support for this work was provided by NASA through the NASA Hubble Fellowship grant \#HST-HF2-51436.001-A awarded by the Space Telescope Science Institute, which is operated by the Association of Universities for Research in Astronomy, Incorporated, under NASA contract NAS5-26555.
\bibliography{report} 
\bibliographystyle{spiebib} 

\end{document}